\documentclass[9pt]{article}
\usepackage{subeqnarray,subfigure}
\usepackage{epsfig,amsmath,amssymb,mathtools}
\usepackage{dirtytalk,enumitem}
\usepackage{mathrsfs,authblk}
\usepackage{tikz,lipsum,lmodern}
\usepackage[most]{tcolorbox}

\usepackage{empheq}
\usepackage{soul,ulem}
\usepackage[pagebackref=true, colorlinks=true]{hyperref}
\definecolor{redish}{rgb}{0.7,0.2,0.0}  
\definecolor{bluish}{rgb}{0.2,0.5,0.8}
\hypersetup{linkcolor=redish,          
	citecolor=blue,        
	filecolor=magenta,      
	urlcolor=bluish}          
\usepackage{setspace}

\title{Critical orbits in Schwarzschild black hole with quintessence and string cloud background space-times}
\author[]{Surya Shankar R\footnote{suryashankar3742@gmail.com}}
\affil[]{{\small Sardar Vallabhbhawe National Institute of Technology, \linebreak Ichchhanath, Surat - Dumas Road,\linebreak Surat 395007, Gujarat, India.  }}
\date{~}               
\begin{document}
\maketitle
\begin{abstract}
In this paper, we have examined the geodesic structure of the spacetime surrounding Schwarzschild black holes in the presence of a quintessence field parameter $q$ and string cloud parameter $a$. Because of the quintessence field, there are two horizon structures and coordinate time $t$ shows some fascinating variations due to the parameters. We do a thorough investigation of the existence of critical orbits for various parameter values.

\end{abstract}
\section{Introduction}
Einstein’s theory of General Relativity \cite{0} has been successful in explaining various observational facts such as gravitational redshift, Mercury's orbital precession, light bending, and so on, as well as predicting various other phenomena such as satellite time-dilation, black hole existence, event horizons, and gravitational collapse. Throughout the last century, astrophysics has been built around the theory, and although the theory is considered to be incomplete in its explanations at sufficiently smaller length scales, it has been widely regarded as one of the most successful theories of all time, and none of its predictions have yet been contradicted. The first solution to Einstein’s equation was the Schwarzschild black hole \cite{1}, and ever since its inception, black holes have been a feature of intrigue for physicists. The study of particle dynamics and photon orbits in black hole spacetimes is crucial from an astronomical standpoint. This research helps us understand the powerful gravitational field that exists around black holes, which in turn helps us understand the deepest secrets of our universe. The recent discovery of gravitational waves emitted by binary black hole mergers has piqued our interest in studying black hole dynamics \cite{2}.\\
The currently accepted cosmological picture of the universe is the Lambda-CDM model \cite{3}. It explains the stability of galaxies and expanding universe with the help of dark matter and dark energy. Dark matter is a hypothetical form of matter thought to account for approximately $85\%$ of the matter in the universe and it does not appear to interact with the electromagnetic spectrum and is therefore difficult to detect. Dark energy is an unknown form of energy that affects the universe on the largest scales and the best current measurements indicate that dark energy contributes $68\%$ of the total energy in the present-day observable universe. Measurements of supernovae provided the first observational evidence for its existence, demonstrating that the cosmos does not expand at a steady rate, but rather accelerates \cite{4}. Quintessence is a hypothesized kind of dark energy, or more accurately, a scalar field proposed as a possible explanation for the universe's rapid expansion rate. The first example of this scenario was proposed by Ratra and Peebles (1988) \cite{5}, and since then, the quintessence field has been incorporated into space-time metrics for more accurate studies.\\
String clouds \cite{6}, can be described as a collection of strings created in the early phases of our Universe due to symmetry breaking. They are predicted by string theory. String clouds have lately been studied in relation to the null and timelike geodesics surrounding the Schwarzschild black hole. A string cloud background makes a profound influence on horizon structure, geodesics and thermodynamic quantities but entropy is not changed. The famous $\Lambda$ constant, which is added to Einstein's field equation can be considered as the first step towards understanding dark matter and dark energy, and it had its own implications for the Schwarzschild metric. A quintessence field can be considered as a more simplistic approach towards incorporating dark energy into the metric of space-time. Blackhole solutions with string cloud background and quintessence field is not only a solution to Einstein’s field equation, but an association of two fundamental theories of theoretical high energy physics and cosmology, with both parameters rendering unique properties to the surrounding space-time.\\
Yet another intriguing aspect of investigating Schwarzschild black hole with quintessence and string clouds, is the difference in the geodesic structures from that of the usual Schwarzschild black hole. The radius of the circular orbits in the Schwarzschild black hole in the presence of the string cloud parameter are larger than the radius of the circular orbits in the Schwarzschild black hole in the absence of the string cloud parameter. Furthermore, it has been discovered that as the value of the string cloud parameter is increased, the particle can more easily escape to infinity. The effect of the quintessence field on the motion of particles in the spacetime of the Schwarzschild black hole has been studied in another recent paper, where it is shown that the quintessence field pushes the circular orbits away from the central object, that is, the radius of the circular orbits are larger than the radius of the circular orbits in the usual Schwarzschild case. As mentioned before, the horizon structures of these 2 black holes are completely different. Most importantly the existence of stable and unstable critical orbits, is a phenomenon worth investigating in itself for Schwarzschild black holes with quintessence and string clouds.\\
Other than the fact that null and timelike geodesics help us understand the structure of the space-time surrounding a black hole, the motion of particles helps us understand the gravitational fields of black holes experimentally and gives us grounds to compare them with observational data. Considering the fact that information from black holes does not reach us because of the event horizon, the study of the inner structures of black holes can only be theoretical. With gravitational waves being incredibly difficult to observe, geodesics studies are the only feasible option available to physicists. Therefore Iinvestigate null and timelike geodesic motion in the vicinity of the Schwarzschild black hole in the presence of the string cloud parameter a and the quintessence field parameter q. The ranges for both the parameters a and q are also determined, which allows the existence of the black hole. Further, the existence of stable and unstable critical orbits around Schwarzschild black holes with quintessence and string clouds are examined.
\section{Horizon strucutre, lagrangian and conserved Quantities}
In the background of quintessence and cloud of strings, the spherically symmetric and static spacetime is represented as
\begin{equation}
 d s^{2}=-\left(1-a-\frac{2 M}{r}-\frac{q}{r^{3 \omega_{q}+1}}\right) d t^{2} +\left(1-a-\frac{2 M}{r}-\frac{q}{r^{3 \omega_{q}+1}}\right)^{-1} d r^{2} +r^{2}\left(d \theta^{2}+\sin ^{2} \theta d \phi^{2}\right)    \label{eq.1}
\end{equation}
where $\omega_{q}, a$, $M$ and $q$ are the equation of state parameter (EoS) for quintessence field, string cloud parameter, mass of the black hole, and quintessence parameter respectively. The Equation of State parameter for the quintessence field has the values, i.e., $-1<\omega_{q}<-\frac{1}{3}$. The current research is limited to a single case when $\omega_{q}=\frac{-2}{3}$. In the absence of $a$ and $q$, the above spacetime can be reduced to the Schwarzschild spacetime. The spacetime metric is,
\begin{equation}
 d s^{2}=-\left(1-a-\frac{2 M}{r}-\frac{q}{r^{-1}}\right) d t^{2} +\left(1-a-\frac{2 M}{r}-\frac{q}{r^{-1}}\right)^{-1} d r^{2} +r^{2}\left(d \theta^{2}+\sin ^{2} \theta d \phi^{2}\right)   \label{eq.2}
\end{equation}
The horizon structure of the spacetime given by eq.(\ref{eq.2}) for $\omega_{q}=\frac{-2}{3}$, is given as
\begin{equation}
   r_{q}=\frac{1-a+\sqrt{a^{2}-2 a-8 M q+1}}{2 q} \quad\quad
   r_{e}=\frac{1-a-\sqrt{a^{2}-2 a-8 M q+1}}{2 q}  \label{eq.3}
\end{equation}
For the quantity in the square root, in eq.(\ref{eq.3}), to be positive Ihave $0<q<\frac{a^{2}-2 a+1}{8 M}$ and $0<a<1$. The horizon owing to the quintessence word is represented by $r_{q}$ and corresponds to the cosmological horizon, whereas the other root, represented by $r_{e}$, corresponds to the event horizon. The existence of string clouds and the quintessence field increases the $r_{e}$. The $r_{e}$, in the required limits reduces to the horizon of the Schwarzschild spacetime i.e. when $a \rightarrow 0$ and $q \rightarrow 0, r_{e}=2 M$. Also to depict an example of the horizon structure, for a unit mass black hole with $a=0.1$, $q=0.001$, in the units $G=1=c$,
$$ r_{e}=2.22773 \quad r_{q}=897.772 $$
For further clarification, for the usual unit mass Schwarzschild black hole the horizon radius $r_{e}=2M=2$. The Lagrangian for the metric given by eq.(\ref{eq.2}) is
\begin{equation}
   \mathfrak{L}= g_{\mu v} \dot{x}^{\mu} \dot{x}^{\nu}=-\left(1-a-\frac{2M}{r}-qr\right)\dot{t^{2}}+\frac{\dot{r^2}}{(1-a-\frac{2M}{r}-qr)}+r^{2}\left(\dot{\theta^{2}}+\sin^{2}{\theta}\dot{\phi^{2}}\right) \label{eq.4}
\end{equation}
Where the four-momentum $p^{\mu}=m u^{\mu}=m \dot{x^{\mu}}$. For this Lagrangian, using Noether symmetries and killing vector equations, Ihave the following well-known results for two conserved quantities, the specific energy $\mathcal{E}$ and the specific angular momentum $\mathcal{L}_{z}$ of the neutral particle in the surrounding spacetime.
\begin{equation}
\mathcal{E} =\xi_{(t)}^{\mu} P_{\mu} / m=-\dot{t} \left(1-a-\frac{2M}{r}-qr\right);
\mathcal{L}_{z}  = \xi_{(\phi)}^{\mu} P_{\mu} / m=\dot{\phi} r^{2} \sin ^{2} \theta \label{eq.5}
\end{equation}
Which is derived using $\xi_{(t)}^{\mu} \partial_{\mu}=\partial_{t}$ and $\xi_{(\phi)}^{\mu} \partial_{\mu}=\partial_{\phi}$ respectively.\\
The overdot denotes the derivative with regard to the proper time $\tau$ here and in the following. The square of the specified total angular momentum is the third integral of motion.
\begin{equation}
    \mathcal{L}^{2} = r^{4} \dot{\theta}^{2}+\frac{\mathcal{L}_{z}^{2}}{\sin ^{2} \theta}=r^{2} U_{\perp}^{2}+\frac{\mathcal{L}_{z}^{2}}{\sin ^{2} \theta} \label{eq.6}
\end{equation}
Here Idenoted $U_{\perp}$ as:
\begin{equation}
    U_{\perp} \equiv-r \dot{\theta}_{o} \label{eq.7}
\end{equation}
Using the normalization condition $\boldsymbol{u_{\mu}u^{\mu}}=-\zeta$, where $\zeta=0,1$ is used to define the null and timelike geodesics respectively, we obtain
\begin{equation}
    \dot{r}^{2}=\mathcal{E}^{2}-U, \quad U=(1-a-\frac{2M}{r}-qr)\left(\zeta+\mathcal{L}^{2} / r^{2}\right) \label{eq.8}
\end{equation}
The particle moves in a flat motion. We can consider this to be in the same plane as the equatorial plane. Then $\dot{\theta}=0$ and the effective potential for the radial motion takes the form
\begin{equation}
    U \equiv U_{eff}= (1-a-\frac{2M}{r}-qr)\left(\zeta+\mathcal{L}_{z}^{2} / r^{2}\right)\label{eq.9}
\end{equation}
\section{Nature of effective potential and critical orbits for timelike geodesics of test particles}
Effective potential given in eq.(\ref{eq.9}) for neutral particles moving along the timelike geodesics for $\zeta=1$,
$$ U_{e f f}(r)=\left(1-a-\frac{2 M}{r}-\frac{q}{r^{-1}}\right)\left(1+\frac{\mathcal{L}_{z}^{2}}{r^{2}}\right) $$
And the equation of motion is 
$$ u^{\mu} u_{\mu}=-1 $$

\subsection{Radial motion of test particle}
\begin{figure}[hbt]
    \begin{center}
		\includegraphics[scale=0.465]{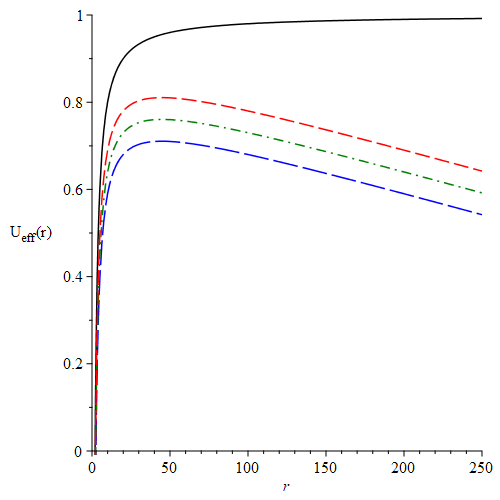}
		\caption{\label{1}Effective potential of a black hole with $M=1$ for radial timelike motion for same value of $q=0.001$ and different values of $a$.$a=0.1$(blue longdash)\quad $a=0.15$(green dashdot)\quad $a=0.2$(red shortdash)
		Solid black line represents $U_{eff}$ of Schwarzschild blackhole}
		\end{center}
	\end{figure}

\begin{figure}[hbt]
\centering
		\includegraphics[scale=0.47]{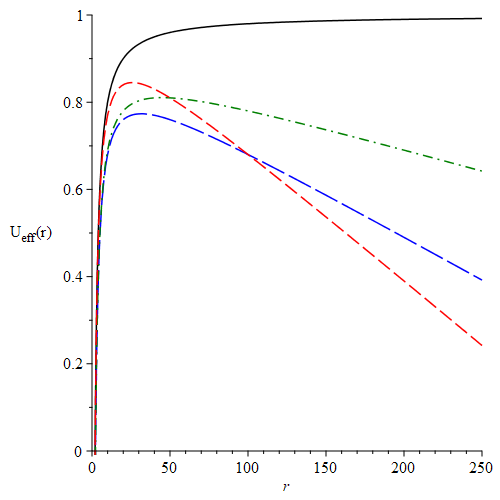}
		\caption{\label{2}Effective potential of a black hole with $M=1$ for radial timelike motion for same value of $a=0.1$ and different values of $q$.$q=0.001$(blue longdash)\quad $q=0.002$(green dashdot)\quad $q=0.003$(red shortdash) Solid black line represents $U_{eff}$ of Schwarzschild blackhole}
	\end{figure}

The effective potential for the given case (when $\mathcal{L}_{z}=0$):
\begin{equation}
    U_{eff} = \left(1 - a - \frac{2M}{r} - qr\right)  \left(1+\frac{0^{2}}{r^{2}}\right) \label{eq.14}
\end{equation}
\begin{equation}
U_{eff} = \left(1 - a - \frac{2M}{r} - qr\right)  \label{eq.15}
\end{equation}
Here Istudy the geodesics for incoming test particles with zero angular momentum. Let us try to understand orbits of the particles in the background of the stringcloud quintessence black holes by using the effective potential $U_{eff}$ given by eq.(\ref{eq.15}). The behaviour of effective potential is shown in Figures (\ref{1}) and (\ref{2}) for a particular set of parameters.

\subsection{Non-radial motion of test particle}
\begin{figure}[hbt]
\begin{center}
		\includegraphics[scale=0.455]{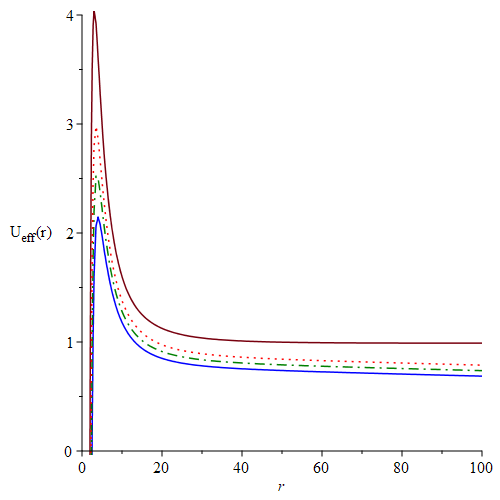}
		\caption{\label{5}Effective potential of a black hole with $M=1$ for non-radial timelike motion with $\mathcal{L}_{z}=10$ for same value of $q=0.001$ and different values of $a$. $a=0.1$(blue longdash),\quad $a=0.15$(green dashdot),\quad $a=0.2$(red shortdash),solid black line represents $U_{eff}$ of Schwarzschild blackhole} 
		\end{center}
	\end{figure}

\begin{figure}[hbt]
\begin{center}
		\includegraphics[scale=0.455]{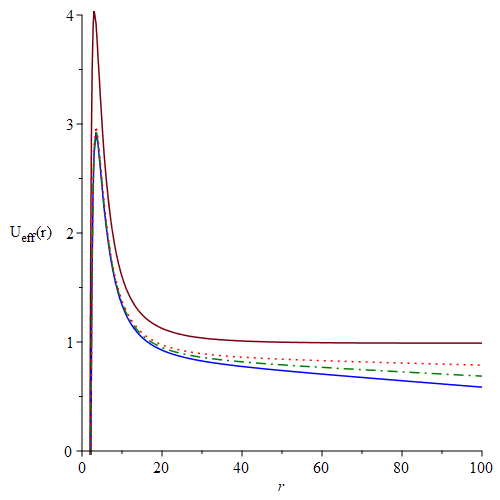}
		\caption{\label{6}Effective potential of a black hole with $M=1$ for non-radial timelike motion with $\mathcal{L}_{z}=10$ for same value of $a=0.1$ and different values of $q$. $q=0.001$(blue longdash),\quad $q=0.002$(green dashdot),\quad $q=0.003$(red shortdash),Solid black line represents $U_{eff}$ of Schwarzschild blackhole}
		\end{center}
	\end{figure}

\begin{figure}[hbt]
\begin{center}
		\includegraphics[scale=0.5]{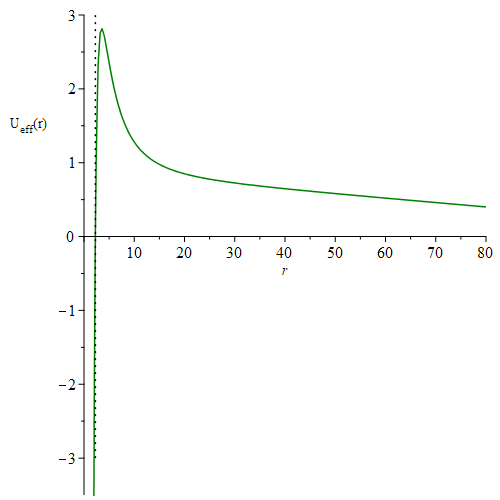}
		\caption{ \label{7}Effective potential of a black hole with $M=1$ for non-radial timelike motion for a non-zero mass particle with $\mathcal{L}_{z}=10$ for $a=0.1$ and $q=0.006$, with horizon radius $r_{e}=2.2561$ depicted using dotted lines, where, as expected $U_{eff}=0$}
		\end{center}
	\end{figure}

\begin{figure}[hbt]
   \begin{center}
       	\includegraphics[scale=0.51]{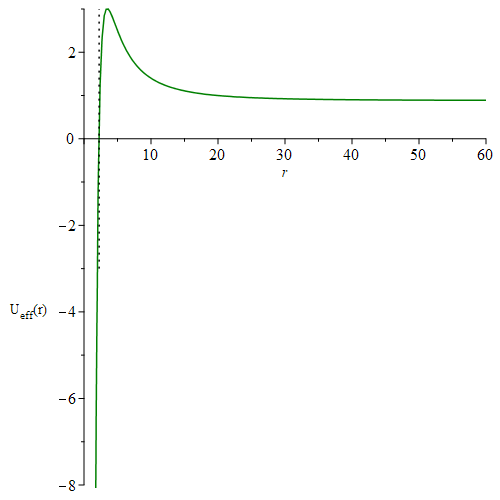}
		\caption{\label{9}Effective potential of a black hole with $M=1$ for Non-Radial Timelike motion for a non-zero mass particle with $\mathcal{L}_{z}=10$ for $a=0.1$ and $q=0.00003$, with horizon radius $r_{e}=2.2561$}
   \end{center}
	\end{figure}

\begin{figure}[hbt]
\begin{center}
    \includegraphics[scale=0.4]{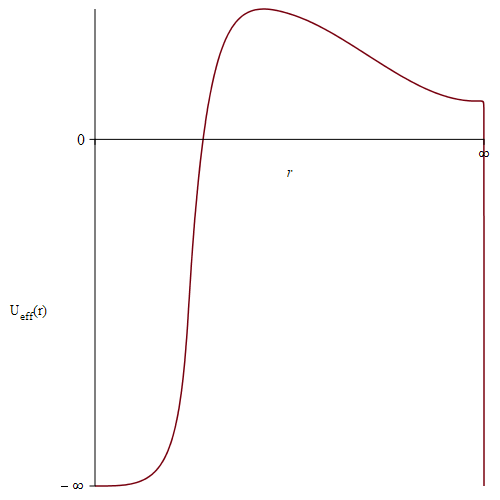}
		\caption{\label{10}Effective potential of a black hole with $M=1$ for Non-Radial Timelike motion for a non-zero mass particle with $\mathcal{L}_{z}=10$ for $a=0.1$ and $q=0.00003$, with horizon radius $r_{e}=2.2561$}
\end{center}
	\end{figure}

The Effective Potential for the given case (when $\mathcal{L}_{z} \neq 0$):
\begin{equation}
  U_{eff} = \left(1 - a - \frac{2M}{r} - qr\right) \left(1+\frac{\mathcal{L}_{z}^{2}}{r^{2}} \right) \label{eq.19}
\end{equation}
Here we study the timelike geodesics for incoming test particles with nonzero angular momentum. Let us try to understand orbits of the particles in the background of the stringcloud quintessence black holes by using the effective potential $U_{eff}$ given by eq.(\ref{eq.19}). Analysing the effective potential, we can see that the shape of the graph and observer orbits are influenced heavily by the parameter quintessence parameter $q$ with variation, whereas the string cloud parameter $a$ being a constant in the metric, must be sufficiently small so as the metric makes as sense. But other than that, the variation of $a$ does not render any peculiar properties to the spacetime surrounding the black hole, and therefore for the time being we can approximate $a$ to a sufficiently small value, say $a=0.1$. Variation of the parameter $q$ is however, important to the existence of stable orbits. Let us analyze the effective potential in detail. The behaviour of effective potential is shown in figures (\ref{5}) and (\ref{6}) for a particular set of parameters.
From the graph it is evident that there exists an unstable critical orbit, but an innermost stable critical orbit is not visible, even for higher values of $r$, as depicted in fig.(\ref{7}).
\begin{figure}[hbt]
\begin{center}
		\includegraphics[scale=0.45]{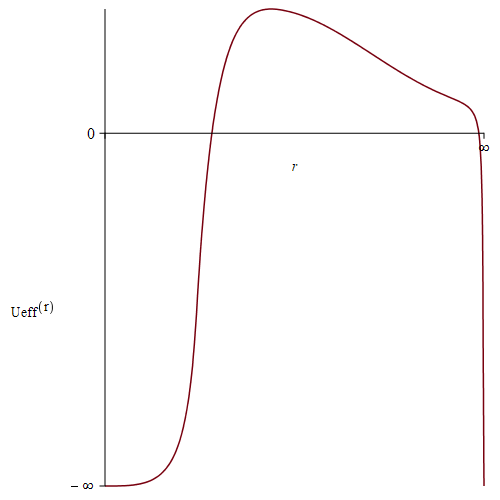}
		\caption{\label{8}Effective potential of a black hole with $M=1$ for Non-Radial Timelike motion for a non-zero mass particle with $\mathcal{L}_{z}=10$ for $a=0.1$ and $q=0.006$, with horizon radius $r_{e}=2.2561$}
		\end{center}
	\end{figure}
Mathematically, the non-existence of stable critical orbit can be proved by obtaining just one real positive solution for the equation $\frac{dU_{eff}}{dr}=0$, for the given set of parameters, and it is well-known that for the usual Schwarzschild metric, $\frac{dU_{eff}}{dr}=0$ produces two real positive solutions, with the conidition satisfied that the observer in motion has a minimum angular momentum $L>2\sqrt{3}M$. In the stringcloud quintessence black hole case,
$$\frac{dU_{eff}}{dr}=  \left( \,{\frac {2M}{{r}^{2}}}-q \right)  \left( 1+{\frac {{\mathcal{L}_{z}}^{2}}{{
r}^{2}}} \right) -2\,{\frac {{\mathcal{L}_{z}}^{2}}{{r}^{3}} \left( 1-a-2\,{\frac {M
}{r}}-qr \right) }
 = 0$$
\begin{equation}
\begin{split}
\implies r= r_{b}^{P}=-\frac{1}{2} \sqrt{-\frac{L^{2}(-q)-2 M}{q}-\frac{L^{2} q+2 M}{3 q}-\frac{\sqrt[3]{2}\left(L^{4} q^{2}-68 L^{2} M q+4 M^{2}\right)}{3 q \sqrt[3]{\Psi_{1}+\Psi_{2}}}-\frac{\sqrt[3]{\Psi_{1}+\Psi_{2}}}{3 \sqrt[3]{2} q}}+\frac{1}{2};\\
 \times \sqrt{-\frac{4\left(a L^{2}-L^{2}\right)}{\Psi_{3}}+\frac{L^{2} q+2 M}{3 q}-\frac{L^{2}(-q)-2 M}{q}+\frac{\sqrt[3]{2}\left(L^{4} q^{2}-68 L^{2} M q+4 M^{2}\right)}{3 q \sqrt[3]{\Psi_{1}+\Psi_{2}}}+\frac{\sqrt[3]{\Psi_{1}+\Psi_{2}}}{3 \sqrt[3]{2} q}} \label{eq.20}
 \end{split}
\end{equation}
\begin{equation}
\begin{split}
\Psi_{1}=\sqrt{\left(-108 q\left(a L^{2}-L^{2}\right)^{2}+2\left(L^{2} q+2 M\right)^{3}+432 L^{2} M q\left(L^{2} q+2 M\right)\right)^{2}-4\left(\left(L^{2} q+2 M\right)^{2}-72 L^{2} M q\right)^{3}} ;\\
\Psi_{2}=-108 q\left(a L^{2}-L^{2}\right)^{2}+2\left(L^{2} q+2 M\right)^{3}+432 L^{2} M q\left(L^{2} q+2 M\right) ;\\
\Psi_{3}=q \sqrt{-\frac{L^{2}(-q)-2 M}{q}-\frac{L^{2} q+2 M}{3 q}-\frac{\sqrt[3]{2}\left(L^{4} q^{2}-68 L^{2} M q+4 M^{2}\right)}{3 q \sqrt[3]{\Psi_{1}+\Psi_{2}}}-\frac{\sqrt[3]{\Psi_{1}+\Psi_{2}}}{3 \sqrt[3]{2} q}} \label{eq.21}
\end{split}
\end{equation}
Considering the given parameters for a unit mass black hole for non-radial timelike motion for a non-zero mass particle with $\mathcal{L}_{z}=10$ for $a=0.1$ and $q=0.006$, with horizon radwe $r_{e}=2.256$
\begin{equation}
    U_{eff}= (1-0.1-\frac{2 \times 1} {r}-0.006r)\left( 1+ \frac{10^{2}}{r^{2}}  \right) \label{eq.22}
\end{equation}
\begin{equation}
   \frac{dU_{eff}}{dr}= \left(\frac{2}{r^{2}}-0.006\right)\left(1+\frac{100}{r^{2}}\right)-\frac{200\left(0.9-\frac{2}{r}-0.006 r\right)}{r^{3}} = 0 \label{eq.23}
\end{equation}
Solutions of this quartic equation
$$ r = 3.505832489,\quad 16.48735813 + 22.58451447 \mathbf{I},\quad -36.48054875,\quad 16.48735813 - 22.58451447 \mathbf{I} $$
The only real positive positive solution is $r= 3.505832489$, which evidently indicates the unstable critical orbit point. This is further supported by the value of the second derivative of $U_{eff}<0$ at the given point, indicating a maxima.
From eq.(\ref{eq.19}),
\begin{equation}
    \frac{d^{2}U_{eff}}{dr^{2}}=-\frac{4\left(1+\frac{100}{r^{2}}\right)}{r^{3}}-\frac{400\left(\frac{2}{r^{2}}-0.006\right)}{r^{3}}+\frac{600\left(0.9-\frac{2}{r}-0.006 r\right)}{r^{4}} \label{eq.24}
\end{equation}
At $r= 3.505832489$, $\frac{d^{2}U_{eff}}{dr^{2}}=-1.077705144$.\\
Therefore we know that for the particular configuration of parameters there exists only an unstable critical point. Analytically we can safely assume that as $q$ approaches zero, with the value of $a$ already sufficiently small, and if the angular momentum is high enough, then the shape of effective potential graph must be close to the Schwarzschild metric case. That is, in addition to an unstable critical orbit, there will also be a stable critical orbit distance. Therefore, we will examine effective potentials for very small values of $q$.
Consider an arbitrary case with $q=0.00003$, a value smaller than any values we have considered till now, we can see that graph (\ref{9}) has gotten sharper and the general graph shape has changed in fig. (\ref{10}). we know
\begin{equation}
  U_{eff}= (1-0.1-\frac{2} {r}-0.00003r)\left( 1+ \frac{10^{2}}{r^{2}}  \right) \label{eq.25}
\end{equation}
Therefore, from eq. (\ref{eq.25})
\begin{equation}
\frac{dU_{eff}}{dr}=\left(\frac{2}{r^{2}}-0.00003\right)\left(1+\frac{100}{r^{2}}\right)-\frac{200\left(0.9-\frac{2}{r}-0.00003 r\right)}{r^{3}}=0 \label{eq.26}
\end{equation}
Real positive solutions $r= 3.467,\quad 103.6568, \quad 188.34$.
$$At \quad r= 3.467, \quad\frac{d^{2}U_{eff}}{dr^{2}}=-1.1496$$
$$At \quad r= 103.6568, \quad \frac{d^{2}U_{eff}}{dr^{2}}=+8.8001 \times10^{-7}$$
$$At \quad r= 188.34, \quad \frac{d^{2}U_{eff}}{dr^{2}}=-1.805889004 \times 10^{-7}$$
Indicating a minima and maxima respectively, while $r= 188.34$ is an anomaly solution similar to the existence of a cosmological horizon, after which the effective potential drops down to $-\infty$. Therefore, for a given configuration of parameters we have proved that for very small values of $q$, both unstable and stable critical orbits exist. This means we also require an investigation as to when the stable critical orbit is lost, for this case and in general. Therefore, we consider the parameter $q$ as a variable. Therefore for unit mass black hole and observer with $\mathcal{L}_{z}=10$,
\begin{equation}
    U_{eff}=(1-0.1-\frac{2\times1}{r}-qr)\left(1+\frac{10^{2}}{r^2}\right) \label{eq.27}
\end{equation}
\begin{equation}
\implies \frac{dU_{eff}}{dr}= \left(\frac{2}{r^{2}}-q\right)\left(1+\frac{100}{r^{2}}\right)-\frac{200\left(0.9-\frac{2}{r}-q r\right)}{r^{3}}=0 \label{eq.28}
\end{equation}
we get a complicated set of solutions and everyone of the solutions has the term,
$$ \left(\sqrt{\frac{50000000 q^{4}+21250000 q^{3}-11025000 q^{2}+1187725 q-46}{q}}\right) $$
Which must be real in-order for the solution to be real, which means that the numerator must be positive, since we already know that only positive values of $q$ are allowed.
\begin{equation}
50000000 q^{4}+21250000 q^{3}-11025000 q^{2}+1187725 q-46  \geq 0 \label{eq.29}
\end{equation}
Considering only real positive solutions for the solution,
\begin{equation}
q = 0.0000387434 \label{eq.30}
\end{equation}
Let us verify our findings
\begin{itemize}
    \item $q = 0.0000387434$
      $$
      U_{eff}= (1-0.1-\frac{2 \times 1} {r}-0.0000387434r)\left( 1+ \frac{10^{2}}{r^{2}}  \right)
      $$
     $$
     \frac{dU_{eff}}{dr}=0 \implies r= 3.467, \quad 130.1,\quad 130.1, \quad -263.7420155
     $$
     Unstable and Stable critical orbits exist.
    \item $q > 0.0000387434$, say $q=0.0000487434$
     $$
      U_{eff}= (1-0.1-\frac{2 \times 1} {r}-0.0000487434r)\left( 1+ \frac{10^{2}}{r^{2}}  \right)
      $$
     $$
     \frac{dU_{eff}}{dr}=0 \implies r= 3.467190995,\quad 117.4693325 + 33.05351446I, \quad -238.4058559,\quad 117.4693325 - 33.05351446I
     $$
     Unstable critical orbit exists while stable critical orbit is lost, as expected.
    \item $q < 0.0000387434$, say $q=0.00003$
     $$
      U_{eff}= (1-0.1-\frac{2 \times 1} {r}-0.00003r)\left( 1+ \frac{10^{2}}{r^{2}}  \right)
      $$
     $$
     \frac{dU_{eff}}{dr}=0 \implies r = 3.4671,\quad 103.656,\quad 188.345,\quad -295.469
     $$
     Unstable and Stable critical orbit exist, as expected.
\end{itemize}
Therefore, we have found that $q=0.0000387434$ is the maximum value of the parameter so that stable circular orbits are not lost for an observer with $\mathcal{L}_{z}=10$ around a unit mass black hole with $a=0.1$.
\section{Generalizing $\mathcal{L}_{zmin}$, $M$ and $q$ for stable and unstable critical orbits to exist}
As we saw from the previous sections, the parameters have to obey certain restrictions for the innermost stable critical orbit (ISCO) to exist and even in general, for any critical orbits at all exist. This is even evident in Schwarzschild black holes where minimum azimuthal momentum $\mathcal{L}_{zmin} = 2\sqrt{3}M$ for both unstable and stable orbits to exist. Due to the increased number of parameters available, we consider 2 variables at a time and then move on to all 3 variables for stringcloud quintessence black holes (assuming that $a=0.1$ in all cases).
\begin{itemize}
    \item \textbf{$\mathcal{L}_{zmin}$ vs $M$ : Consider a black hole with $a=0.1$ and $q=0.00003$.} 
\begin{equation}
    U_{eff}=(1-0.1-\frac{2M}{r}-0.00003r)\left(1+\frac{\mathcal{L}_{z}^{2}}{r^{2}}\right) \label{eq.31}
\end{equation}
\begin{equation}
\frac{dU_{eff}}{dr}=\left(\frac{2 M}{r^{2}}-0.00003\right)\left(1+\frac{\mathcal{L}_{z}^{2}}{r^{2}}\right)-\frac{2\left(0.9-\frac{2 M}{r}-0.00003 r\right) \mathcal{L}_{z}^{2}}{r^{3}}=0 \label{eq.32}
\end{equation}
From the solution of this equation we derive, for real solutions to exist,
\begin{equation}
\begin{split}
    1215 M \mathcal{L}_{z}^{10}-5467500 \mathcal{L}_{z}^{10}+2.268\times 10^{9} M^{2} \mathcal{L}_{z}^{8}- 4.04595\times 10^{13} \mathcal{L}_{z}^{8} M+1.0692\times 10^{15} M^{3} \mathcal{L}_{z}^{6};\\
+1.3286025\times 10^{17} \mathcal{L}_{z}^{8}-2.6973\times 10^{18} \mathcal{L}_{z}^{6} M^{2}+1.008\times 10^{19} M^{4} \mathcal{L}_{z}^{4};
-1.62\times 10^{21} \mathcal{L}_{z}^{4} M^{3}+2.4\times 10^{22} M^{5} \mathcal{L}_{z}^{2} \geq 0 \label{eq.33}
\end{split}
\end{equation}
$\mathcal{L}_{zmin} \approx 3.848$ for a black hole with $M=1$, $q=0.00003$ and $a=0.1$. Due to the linear nature of the graph in figure (\ref{11}), we cannot derive any particular information about existence of stable and unstable critical orbits separately, rather, it gives us the information that for any orbits to exist, the above relation must be satisfied. Specifically, if $\mathcal{L}_{zmin}<3.848$ for a black hole with $M=1$, $q=0.00003$ and $a=0.1$, no orbits exist, similar to the Schwarzschild case (where $\mathcal{L}_{zmin} \approx 3.464$ for $M=1$).
	\begin{figure}[hbt]
\begin{center}
    	\includegraphics[scale=0.6]{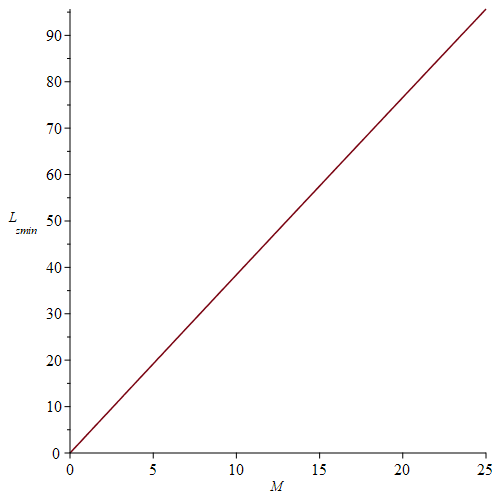}
		\caption{\label{11}$\mathcal{L}_{zmin}$ vs $M$ of a black hole for Non-Radial Timelike observer for $q=0.00003$ and $a=0.1$}
\end{center}
\end{figure}
\begin{center}
\begin{tabular}{ cc }
$\mathcal{L}_{z}$ v/s $r_{C}$ (for $M=1$) & $M$ v/s $r_{C}$ (for $\mathcal{L}_{z}=10$) \\  
\begin{tabular}{ |c|c| } 
\hline
$\mathcal{L}_{z}$ & $r_{C}$ \\
\hline
5 & 4.0694, 18.544, 246.34 \\
10 & 3.467, 103.65, 188.34  \\
15 & 3.390  \\
20 & 3.365  \\
25 & 3.3535 \\
\hline
\end{tabular} &  
\begin{tabular}{ |c|c| } 
\hline
$M$ & $r_{C}$ \\
\hline
1 & 3.467, 103.656, 188.34 \\
2 & 8.139, 37.324, 340.84  \\
3 & 431.89  \\
4 & 505.17 \\
5 & 568.49 \\
\hline
\end{tabular} \\
\end{tabular}
\end{center}
\vspace{20 pt}
\item \textbf{$\mathcal{L}_{z}$ vs $q$ : Consider a unit mass black hole with $a=0.1$.}
\begin{equation}
    U_{eff}=(1-0.1-\frac{2}{r}-qr)\left(1+\frac{\mathcal{L}_{z}^{2}}{r^{2}}\right) \label{eq.34}
\end{equation}
\begin{equation}
    \frac{dU_{eff}}{dr}= \left(\frac{2}{r^{2}}-q\right)\left(1+\frac{\mathcal{L}_{z}^{2}}{r^{2}}\right)-\frac{2\left(0.9-\frac{2}{r}-q r\right) \mathcal{L}_{z}^{2}}{r^{3}} =0 \label{eq.35}
\end{equation}
From the solution of this equation we derive, for real solutions to exist,
\begin{equation}
\begin{split}
    20000 \mathcal{L}_{z}^{8} q^{4}-2700 \mathcal{L}_{z}^{8} q^{3}+1120000 \mathcal{L}_{z}^{6} q^{3}-599400 \mathcal{L}_{z}^{6} q^{2}; \\
+59049 \mathcal{L}_{z}^{6} q+15840000 \mathcal{L}_{z}^{4} q^{2}-1198800 \mathcal{L}_{z}^{4} q+4480000 q \mathcal{L}_{z}^{2}-21600 \mathcal{L}_{z}^{2}+320000 \geq 0 \label{eq.36}
\end{split}
\end{equation}
\begin{figure}[hbt]
\begin{center}
    	\includegraphics[scale=0.6]{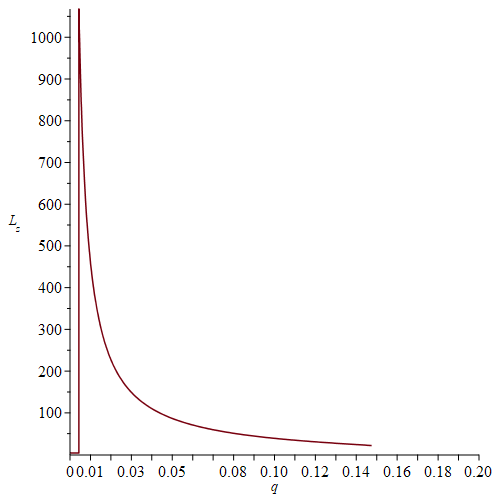}
		\caption{\label{12}$\mathcal{L}_{z}$ vs $q$ of a unit mass black hole for Non-Radial Timelike observer for $a=0.1$}
\end{center}
\end{figure}
From the graph in Fig. (\ref{12}) and even analytically, it is evident that $\mathcal{L}_{z}$ and $q$ values are directly related to the existence of stable and unstable critical orbits. As we already know, the metric itself loses meaning as we use increased values of $q$, which in this case is around $q-0.15$. Due to the shape of the graph we can assume logically that for sufficiently small values of $q$, very large values of $\mathcal{L}_{z}$ are possible to sustain both stable and unstable critical orbits. And as we can see, once the value of $q$ is higher than the peak, we know that stable critical orbits are lost, while unstable critical orbits are possible for much smaller values of $\mathcal{L}_{z}$, which is again being depicted here. This also means that, for a value of $q$ small enough to sustain both stable and unstable critical orbit, there is not only a minimum value, but also a maximum value of $\mathcal{L}_{z}$, above which stable critical orbits are lost. That means within the parameters, the black hole cannot sustain an observer with very high angular momentum in a stable orbit and it would presumably either fly off, or cross the horizon and crash into the singularity. Let us consider the case of the unit mass black hole with $q=0.00003$ and $a=0.1$.
\begin{equation}
    U_{eff}=(1-0.1-\frac{2}{r}-0.00003r)\left(1+\frac{\mathcal{L}_{z}^{2}}{r^{2}}\right) \label{eq.37}
\end{equation}
\begin{equation}
\frac{dU_{eff}}{dr}=\left(\frac{2}{r^{2}}-0.00003\right)\left(1+\frac{\mathcal{L}_{z}^{2}}{r^{2}}\right)-\frac{2\left(0.9-\frac{2}{r}-0.00003 r\right) \mathcal{L}_{z}^{2}}{r^{3}}=0 \label{eq.38}
\end{equation}
For real solutions,
\begin{equation}
\begin{split}
    -5466285 \cdot \mathcal{L}_{z}^{10}+132819792768000000 \cdot \mathcal{L}_{z}^{8}-2696230800000000000 \cdot \mathcal{L}_{z}^{6};\\
-1609920000000000000000 \cdot \mathcal{L}_{z}^{4}+24000000000000000000000 \cdot \mathcal{L}_{z}^{2} \geq 0 \label{eq.39}
\end{split}
\end{equation}
Real positive solutions to the equation,
$$
\mathcal{L}_{z}=3.848,\quad 10.642,\quad 155878.16
$$
Let us examine all 3 solutions.
\begin{itemize}
    \item $\mathcal{L}_{z}=3.848$ : Is the minimum value of $\mathcal{L}_{z}$ for orbits to exist. $\frac{dU_{eff}}{dr}=0$ has two real positive solutions.\\ Any value less than this, say $\mathcal{L}_{z}=3.7$, $\frac{dU_{eff}}{dr}=0$ will have no real positive solutions, meaning no orbits are possible. Any value less than this, say $\mathcal{L}_{z}=3.9$, $\frac{dU_{eff}}{dr}=0$ has 2 real positive solutions, meaning both stable and unstable critical orbits exist, as expected.
    \item $\mathcal{L}_{z}=10.642$ : Is the maximum value of $\mathcal{L}_{z}$ for stable critical orbits to exist. $\frac{dU_{eff}}{dr}=0$ has only one real positive solution.\\
    Any value less than this, say $\mathcal{L}_{z}=10.6$, $\frac{dU_{eff}}{dr}=0$ will have two real positive solutions, meaning both stable and unstable critical orbits exist, as expected. Any value more than this, say $\mathcal{L}_{z}=10.7$, $\frac{dU_{eff}}{dr}=0$ will have only one real positive solution, meaning stable critical orbit is lost.
    \item $\mathcal{L}_{z}=155878.16$ : This is an unusually large number that appears to be an anomaly solution similar to the existence of the cosmological horizon existing at such large distances. This would also explain why the values of $\mathcal{L}_{z}$ in the graph of $\mathcal{L}_{z}$ vs $q$ are unusually high for appropriate values of $q$.
\end{itemize}
\begin{center}
\begin{tabular}{ cc }   
$\mathcal{L}_{z}$ v/s $r_{C}$ (for $q=0.00002$) & $q$ v/s $r_{C}$ (for $\mathcal{L}_{z}=10$) \\  
\begin{tabular}{ |c|c| } 
\hline
$\mathcal{L}_{z}$ & $r_{C}$ \\
\hline
5 & 4.0693, 18.506, 304.49 \\
10 & 3.467, 95.461, 255.64 \\
15 & 3.3902  \\
20 & 3.3649 \\
25 & 3.3534 \\
\hline
\end{tabular} &  
\begin{tabular}{ |c|c| } 
\hline
$q$ & $r_{C}$ \\
\hline
0.00001 & 3.467, 90.316, 393.35\\
0.00003 & 3.467, 103.65, 188.34  \\
0.00005 & 3.4671  \\
0.00007 & 3.4673 \\
0.00009 & 3.4674 \\
\hline
\end{tabular} \\
\end{tabular}
\end{center}
\vspace{25 pt}
\item \textbf{$\mathcal{L}_{zmin}$ vs $q$ vs $M$} : Consider a black hole with $a=0.1$
\end{itemize}
\begin{equation}
    U_{eff}=(1-0.1-\frac{2M}{r}-qr)\left(1+\frac{\mathcal{L}_{z}^{2}}{r^{2}}\right) \label{eq.40}
\end{equation}
\begin{equation}
\frac{dU_{eff}}{dr}=\left(\frac{2 M}{r^{2}}-q\right)\left(1+\frac{\mathcal{L}_{z}^{2}}{r^{2}}\right)-\frac{2\left(0.9-\frac{2 M}{r}-q r\right) \mathcal{L}_{z}^{2}}{r^{3}}=0 \label{eq.41}
\end{equation}
For real solutions to exist,
\begin{equation}
\begin{split}
20000 \mathcal{L}_{z}^{8} M q^{4}-2700 \mathcal{L}_{z}^{8} q^{3}+1120000 \mathcal{L}_{z}^{6} M^{2} q^{3}-599400 \mathcal{L}_{z}^{6} M q^{2}+15840000 \mathcal{L}_{z}^{4} M^{3} q^{2};\\
+59049 \mathcal{L}_{z}^{6} q-1198800 \mathcal{L}_{z}^{4} M^{2} q+4480000 \mathcal{L}_{z}^{2} M^{4} q-21600 \mathcal{L}_{z}^{2} M^{3}+320000 M^{5} \geq 0 \label{eq.42}
\end{split}
\end{equation}
	\begin{figure}[hbt]
\begin{center}
		\includegraphics[scale=1.0]{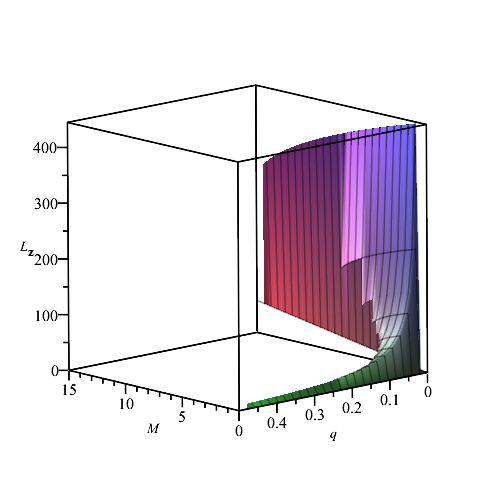}
		\caption{\label{13}$\mathcal{L}_{z}$ vs $M$ vs $q$ of a black hole for Non-Radial Timelike observer for $a=0.1$}
		\end{center}
	\end{figure}
From the graph fig. (\ref{13}), and from previous assumptions we know that only values of $q$ very close to 0 are possible, which is again being shown here, as possible values of angular momentum decrease rapidly as $q$ increases, and the metric loses meaning as $q$ approaches $0.5$. On the other hand as $q$ approaches zero the metric becomes more stable and we can see that a linear relation forms between $M$ and $\mathcal{L}_{z}$ as expected.
\section{Nature of effective potential and classification of orbits for null geodesics of massless particles and photons}
Effective potential given in eq.(\ref{eq.9})  for particles moving along the null geodesics for $\zeta=0$,
$$ U_{eff}(r)=\left(1-a-\frac{2 M}{r}-\frac{q}{r^{-1}}\right)\frac{\mathcal{L}_{z}^{2}}{r^{2}} $$
And the equation of motion is 
$$ u^{\mu} u_{\mu}=0 $$
\subsection{Radial motion}
Since the radial geodesics are the trajectories followed by zero angular momentum test particles ($\mathcal{L}_{z}=0$ ) in given spacetime geometry, the effective potential given for radial null geodesics vanishes i.e. $U_{eff}=0$ and consequently the radial equation of motion reduces to, 
\begin{equation}
  \frac{d r}{d \tau}=\pm \mathcal{E}   \label{eq.43}
\end{equation}
where positive sign corresponds to the outgoing test particles. The integration of this equation leads to,
\begin{equation}
    r=\pm \mathcal{E} \tau+\tau_{0} \label{eq.44}
\end{equation}
where $\tau_{0}$ is the integration constant corresponding to the initial position of the test particle. Shows that in terms of proper time, the radial coordinate depends only on the constant energy value $\mathcal{E}$.
\begin{equation}
    \mathcal{E}=-\dot{t}(1-a-\frac{2M}{r}-qr) \label{eq.45}
\end{equation}
From equations (\ref{eq.43}) and (\ref{eq.45}),
$$
\left(\frac{\dot{r}}{\dot{t}}\right)^{2}=\frac{\mathcal{E}^{2}}{\mathcal{E}^{2}(1-a-\frac{2M}{r}-qr)^{-2}}
$$
\begin{equation}
  \implies \left(\frac{d r}{d t}\right)=\pm\left(1-a-\frac{2 M}{r}-\frac{q}{r^{-1}}\right) \label{eq.46}
\end{equation}
Integrating eq.(\ref{eq.46}) we arrive at the expression for the coordinate time $t$ for the outgoing and ingoing photons as
\begin{equation}
  t=\frac{\mp \log (r(a+q r-1)+2 M) \pm \frac{2(a-1) \tan ^{-1}\left(\frac{a+2 q r-1}{\sqrt{8 M q-(a-1)^{2}}}\right)}{\sqrt{8 M q-(a-1)^{2}}}}{2 q}   \label{eq.47}
\end{equation}

It can be observed that as the values of both the parameters $a$ and $q$ grow, the time $t$ decreases, and as the photon approaches the black hole's horizon, a turning point appears, after which the influence of the quintessence field is exactly opposite on the time $t$. It indicates that in the presence of string clouds and quintessence, the particle moves faster towards the black hole for an external observer before slowing down under the effect of the quintessence field after reaching the turning point. When compared to a pure Schwarzschild black hole, this finding demonstrates that the particle takes less time to reach the horizon.

\subsection{Non-radial motion}
\begin{figure}[hbt]
\begin{center}
    \includegraphics[scale=0.45]{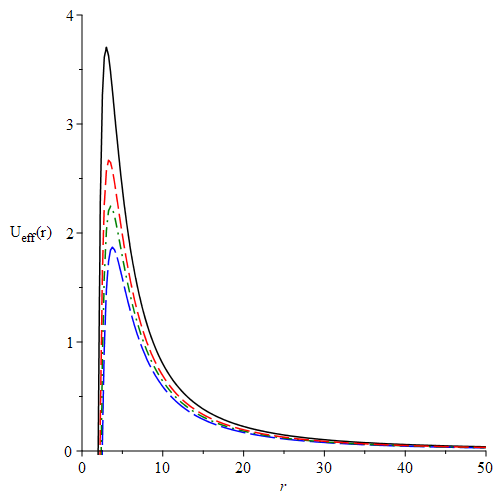}
		\caption{\label{18}Effective potential of a black hole with $M=1$ for non-radial null motion with $\mathcal{L}_{z}=10$ for same value of $q=0.001$ and different values of $a$. $a=0.1$(blue longdash)\quad $a=0.15$(green dashdot)\quad $a=0.2$(red shortdash). Solid black line represents $U_{eff}$ of Schwarzschild blackhole}
\end{center}
\end{figure}

\begin{figure}[hbt!]
\begin{center}
    	\includegraphics[scale=0.45]{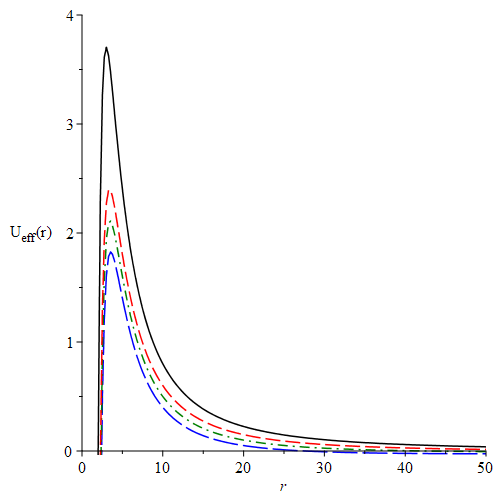}
		\caption{\label{19}Effective potential of a black hole with $M=1$ for non-radial null motion with $\mathcal{L}_{z}=10$ for same value of $a=0.1$ and different values of $q$. $q=0.01$ (blue longdash)\quad $q=0.02$(green dashdot)\quad $q=0.03$(red shortdash). Solid black line represents $U_{eff}$ of Schwarzschild blackhole}
\end{center}
\end{figure}

\begin{equation}
    U_{e f f}(r)=\left(1-a-\frac{2 M}{r}-qr\right)\frac{\mathcal{L}_{z}^{2}}{r^{2}} \label{eq.48}
\end{equation}
The radii of the unstable photon orbit can be derived by taking
$\frac{dU_{eff}}{dr} = 0$ we get
$$\implies \left(\frac{2 M}{r^{2}}-q\right)\frac{\mathcal{L}_{z}^{2}}{r^{2}} - \left(1-a-\frac{2 M}{r}-qr\right)\frac{2\mathcal{L}_{z}^{2}}{r^{3}}=0 $$
\begin{equation}
r_{q}^{N}=\frac{1-a-\sqrt{a^{2}-2 a-6 M q+1}}{q}; 
r_{e}^{N}=\frac{1-a+\sqrt{a^{2}-2 a-6 M q+1}}{q} \label{eq.49}
\end{equation}
In the required limit $r_{e}^{N}$, reduces to that for the Schwarzschild case when $a=q=0$. The $r_{q}^{N}$, is due to the presence of the quintessence field. Only an unstable critical orbit exists and no stable critical orbit.\\
\begin{figure}[hbt]
\begin{center}
    \includegraphics[scale=0.5]{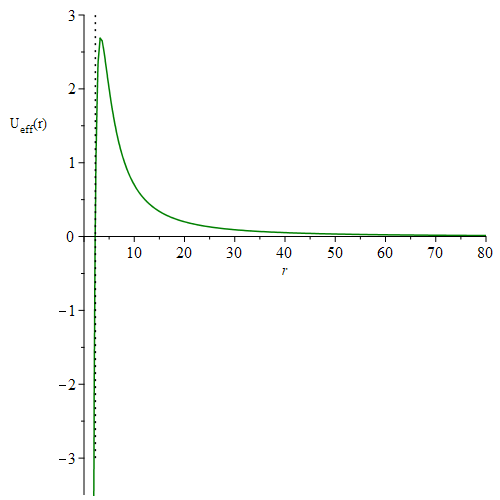}
		\caption{\label{16}Effective potential of a black hole with $M=1$ for non-radial null geodesic motion for massless particle with $\mathcal{L}_{z}=10$ for $q=0.00003$ and $a=0.1$, with horizon radius $r_{e}=2.2561$}
\end{center}
\end{figure}
\begin{figure}[htb!]
\begin{center}
    \includegraphics[scale=0.43]{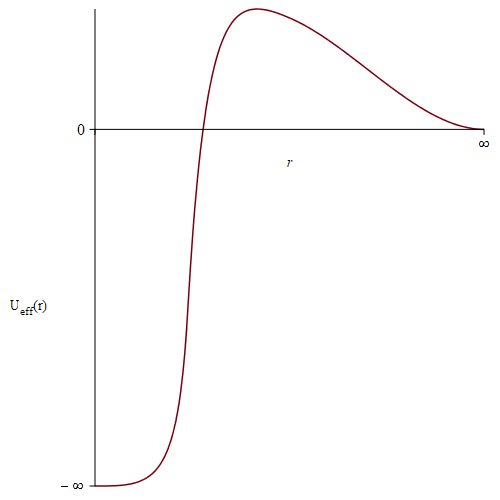}
		\caption{\label{17}Effective potential of a black hole with $M=1$ for non-radial null geodesic motion for massless particle with $\mathcal{L}_{z}=10$ for $q=0.00003$ and $a=0.1$, with horizon radius $r_{e}=2.2561$}
\end{center}
\end{figure}
The general behaviour of effective potential for massless particles is shown in figures (\ref{16}) and (\ref{17}), and behaviour of effective potential is shown in figures (\ref{18}) and (\ref{19}) for a particular set of parameters
\pagebreak
\newpage
\subsubsection{Circular Orbit}
    From eq.(\ref{eq.49}), we know that at $r=r_{e}^{N}=\frac{1-a+\sqrt{a^{2}-2 a-6 M q+1}}{q}$, when massless particle is at $B$. For circular orbits, $r=r_{e}^{N}$ = constant, where $r_{e}^{N}$ is the radius of the circular orbit from the singularity and hence $\dot{r} = 0$. We may use the result to compute the time periods for circular orbits, $\Delta t= T_{t} $ and $\Delta \phi=2\pi$ to obtain
\begin{equation}
    \frac{dt}{d\phi}=r_{e}^{N}\left(1-a-\frac{2M}{r_{e}^{N}}-qr_{e}^{N}\right)^{\frac{-1}{2}} \label{eq.50}
\end{equation}
\begin{equation}
    \therefore T_{t}= 2\pi r_{e}^{N}\left(1-a-\frac{2M}{r_{e}^{N}}-qr_{e}^{N}\right)^{\frac{-1}{2}} \label{eq.51}
\end{equation}
From the effective potential figures it can be very clearly seen that, for photons, in all cases there exists just one critical unstable radius or the photon sphere, whose radius is $r^{N}_{e}$. Stable critical orbits do not exist whatsoever.
\section{Conclusion}
With string clouds and the quintessence scalar field background, we examined the dynamics of particles and photons for Schwarzschild black hole in this paper. Effective potential of black hole generally increases with increase in value of parameter $a$, while increasing value of parameter $q$ results in higher peak for effective potential, for non-massless observers. in both radial and non-radial motion. For timelike obsevers, we have observed how the shape of the effective potential graph changes with increase in value of $q$ parameter and how stable critical orbits disappear after a certain value of $q$. When we increased the values of the parameters $a$ and $q$ in the null particle example, we noticed that the coordinate time t decreased at first, then increased after reaching a turning point as the parameter $q$ increased.\\
Further, the radial geodesics of null and timelike particles for the feasible ranges of the parameters $a$ and $q$ have been addressed. The value of $q$ parameter has to be significantly smaller compared to $a$ parameter for stable critical orbits. Implying the quintessence parameter should be negligible, compared to the stringcloud parameter for these black holes to behave similar to the usual Schwarzschild black hole. Nevertheless, the existence of a quintessence field means, at infinity, the effective potential drops to negative $-\infty$ for timelike observers which seems unfeasible. Which can be translated to a mathematical abberation specifying that the quintessence field should almost be negligible for black holes. Effective potential for null observers however drops down to zero at infinity.\\
For timelike obsevers, minimum angular momentum required and mass of black hole form a linear relation similar to that of Schwarzschild black hole. Because of the graph's linear nature, we can't deduce any specific information regarding the existence of stable and unstable critical orbits individually; instead, it tells us that the linear minimum angular momentum versus mass relation must be satisfied for any orbit to exist.\\ 
We also observed that in the presence of stringcloud and quintessence fields, for a given configuration of $a$ and $q$, we also have a minimum and maximum value of angular momentum beyond which stable orbits cannot be sustained. We can deduce from the graph's form that for sufficiently tiny quantities of the quintessence field parameter, very large values of angular momentum can support both stable and unstable critical orbits. As we can see, once the quintessence field parameter exceeds the peak, stable critical orbits are lost, although unstable critical orbits are conceivable for much lower angular momentum values, as seen above. This also suggests that there is not only a minimum but also a maximum value of angular momentum for a value of quintessence field parameter small enough to maintain both stable and unstable critical orbits.\\
Nature of effective potential for null observers is already very similar to that of the usual Schwarzschild black hole and therefore a photon sphere exists in the geodesic structure. We have therefore concluded that the metric is effectively meaningless as we consider values of $a$ and $q$ beyond minute values and the linearity relation for minimum angular momentum and mass only exist if the metric is effective. When we examine the effective potential, we can see that the parameter quintessence parameter $q$ with variation has a significant influence on the shape of the graph and observer orbits, whereas the string cloud parameter $a$, which is a constant in the metric, must be small enough for the metric to make sense. However, because the variation of $a$ lends no distinctive features in the spacetime around the black hole, we can estimate $a$ to a sufficiently tiny value. The presence of stable orbits depends on the variation of the parameter $q$. In essence, when the configuration of the stringcloud and quintessence parameter is feasible, the black hole behaves quiet similar to the usual Schwarzschild black hole.\\

\end{document}